\documentclass{PoS}
\usepackage{amsmath}
\usepackage{booktabs}
\usepackage{amssymb}
\usepackage{latexsym}
\usepackage{wrapfig}
\usepackage{subfigure}

\title{Way to crosscheck $\mu$-$e$ conversion in the case of no signals of 
$\mu \to e \gamma$ and $\mu \to 3e$}

\ShortTitle{Way to crosscheck $\mu$-$e$ conversion in the case of no signals of 
$\mu \to e \gamma$ and $\mu \to 3e$}

\author{\speaker{Masato Yamanaka}\\
        Kobayashi-Maskawa Institute for the Origin of Particles and
 the Universe (KMI),\\
 Nagoya University, Nagoya 464-8602, Japan\\
        E-mail: \email{yamanaka@eken.phys.nagoya-u.ac.jp}}

\abstract{We consider the case that $\mu$-$e$ conversion 
signal is discovered but other charged lepton flavor violating 
(cLFV) processes will never be found. In such a case, we need 
other approaches to confirm the $\mu$-$e$ conversion and 
its underlying physics without conventional cLFV searches. 
We study R-parity violating (RPV) SUSY models as a benchmark. 
We briefly review that our interesting case is realized in 
RPV SUSY models with reasonable settings according to current 
theoretical/experimental status. 
We focus on the exotic collider signatures at the LHC ($pp \to 
\mu^- e^+$ and $pp \to jj$) as the other approaches. We 
show the correlations between the branching ratio of $\mu$-$e$ 
conversion process and cross sections of these processes. 
It is first time that the correlations are graphically shown. 
We exhibit the RPV parameter dependence of the branching 
ratio and the cross sections, and discuss the feasibility to 
determine the parameters. 
This paper is based on Ref.~\cite{RPV}.}

\FullConference{Flavor Physics \& CP Violation 2015,\\
		May 25-29, 2015\\
		Nagoya, Japan}

\begin{document} 

~

\vspace{-19mm}
\section{Introduction}
\vspace{-3.5mm}

Lepton flavor violation (LFV) is the clearest signal for physics beyond the Standard 
Model (SM)~\cite{Kuno:1999jp}, and searches for LFV have been made 
\cite{Brooks:1999pu, Adam:2013mnn, Bertl:2006up, Bellgardt:1987du}.  
LFV had been found in neutrino oscillation and it indeed requires us to 
extend the SM so that physics beyond the SM includes LFV.  This gives us a 
strong motivation to search for charged lepton flavor violation (cLFV). 
New experiments, COMET~\cite{Cui:2009zz, Kuno:2013mha} and 
DeeMe \cite{Natori:2014yba}, will launch soon and search 
$\mu$-$e$ conversion.  
If COMET/DeeMe observe the $\mu$-$e$ conversion, then with what
kind of new physics should we interpret it?  Now it is worth 
considering since we are in-between two kinds of cLFV experiments
with muon.

For these several decades,  supersymmetric (SUSY) theories have 
been most studied. These include a source 
of LFV.  In the theories, with the R-parity conservation, $\mu \to e 
\gamma$ has the largest branching ratio among the muon cLFV 
\cite{Hisano:1995cp,Sato:2000ff,Koike:2010xr}. 
This occurs via the dipole operator and the other two, $\mu-e$ 
conversion and $\mu \to 3e$, are realized by attaching a quark and 
an electron line at the end of the photon line respectively, giving 
an $\mathcal{O}$($\alpha$) suppression.  
Those branching ratios must be smaller than that of $\mu \to e\gamma$.  
At this moment, the bounds for the branching ratios 
are almost same each other. It means if COMET/DeeMe observe the 
$\mu-e$ conversion, we have to discard  this scenario.

It is, however, possible to find a theory in which COMET/DeeMe find cLFV 
first. The $\mu \to e \gamma$ occurs only 
at loop level due to the gauge invariance, while other two can occur as a 
tree process. Therefore in this case we have to consider a theory in which 
the $\mu-e$ conversion occurs as tree process.
So we have to assume a particle which violate muon and electron number. 
Since $\mu - e$ conversion occurs in a nucleus, it also couples with quarks 
with flavor conservation. 
Furthermore it is better to assume that it does not couple with two 
electrons as we have not observed $\mu\rightarrow 3e$.

We consider the case that COMET/DeeMe indeed observe the cLFV, 
while all the other experiments will not observe anything new.  
In this case other new physics signals are expected to be
quite few, since the magnitude of the cLFV interaction is so small due
to its tiny branching ratio. Therefore it is very important to simulate
now how to confirm the COMET signal and the new physics.  As a benchmark
case we study SUSY models without R parity.

\vspace{-5mm}
\section{RPV Interaction and Our Scenario}  \label{Sec:interaction} 
\vspace{-3.3mm}

The gauge invariant superpotential contains 
the R-parity violating terms~\cite{Weinberg:1981wj, Sakai:1981pk, 
Hall:1983id}, $\mathcal{W}_\text{RPV} = \lambda_{ijk} L_i L_j E_k^c 
   + \lambda'_{ijk} L_i Q_j D_k^c 
   + \lambda''_{ijk} U_i^c D_j^c D_k^c$. 
Here $E_i^c$, $U_i^c$ and $D_i^c$ are $SU(2)_L$ singlet, and $L_i$ 
and $Q_i$ are $SU(2)_L$ doublet superfields. Indices $i$, $j$, 
and $k$ represent the generations. 
We take $\lambda_{ijk} = - \lambda_{jik}$ and $\lambda''_{ijk} 
= - \lambda''_{ikj}$. First two terms include lepton number violation, and 
the last term includes baryon number violation. Since combinations 
of them accelerate proton decay, we omit the last term.

Our interesting situation is that only $\mu$-$e$ conversion is discovered,
and other cLFV processes will not be observed. The situation is realized 
under the following 3 settings on the RPV interaction: 
\begin{enumerate}
\vspace{-2mm}
\item 
only the third generation slepton contributes to the RPV interactions 
\vspace{-3mm}
\item
for quarks, flavor diagonal components are much larger than that 
of off-diagonal components, i.e., CKM-like matrix, $\lambda'_{ijj} 
\gg \lambda'_{ijk} (j \neq k)$
\vspace{-3mm}
\item 
generation between left-handed and right-handed leptons are 
different, $\lambda_{ijk} (i \neq k \text{ and } j \neq k)$. 
\end{enumerate}
\vspace{-2mm}
The setting-1 is realized by the RG evolved SUSY spectrum with 
universal soft masses at the GUT scale. For the simplicity, we decouple 
SUSY particles except for the third generation sleptons. 
The setting-2 is also realized in most cases unless we introduce extra 
sources of flavor violations. 
The setting-3 is artificially introduced to realize the interesting situation.
As a result, the Lagrangian is 
\begin{equation}
\begin{split}
   &
   \mathcal{L}_\text{RPV} 
   =  2 \bigl[ 
   \lambda_{312} \tilde \nu_{\tau L} \overline{\mu} P_L e 
   + \lambda_{321} \tilde \nu_{\tau L} \overline{e} P_L \mu 
   + \lambda_{132} \tilde \tau_L \overline{\mu} P_L \nu_e 
   + \lambda_{231} \tilde \tau_L \overline{e} P_L \nu_\mu
   \\& \hspace{10mm}  
   + \lambda_{123} \tilde \tau_R^* \overline{(\nu_{eL})^c} P_L \mu 
   + \lambda_{213} \tilde \tau_R^* \overline{(\nu_{\mu L})^c} P_L e
   \bigr] + \text{h.c.}, 
   \\[0mm]& \hspace{10mm}   
   + \bigl[ 
   \lambda'_{311} \bigl( \tilde \nu_{\tau L} \overline{d} P_L d 
   - \tilde \tau_L \overline{d} P_L u \bigr) 
   + \lambda'_{322} \bigl( \tilde \nu_{\tau L} \overline{s} P_L s 
   - \tilde \tau_L \overline{s} P_L c \bigr)
   \bigr] + \text{h.c.}.  
\label{Eq:RPV_L2}   
\end{split}      
\end{equation}

Processes described by the Lagrangian \eqref{Eq:RPV_L2} 
strongly depend on the values of $\lambda'_{311}$ and $\lambda'_{322}$. 
To clarify the dependence and to discuss the discrimination of each 
other, we study three cases~\cite{RPV}: [case-I] 
$\lambda'_{311} \neq 0$ and $\lambda'_{322} = 0$, 
[case-I\hspace{-1pt}I] $\lambda'_{311} = 0$ and $\lambda'_{322} \neq 0$, 
and [case-I\hspace{-1pt}I\hspace{-1pt}I]  $\lambda'_{311} \neq 0$ 
and $\lambda'_{322} \neq 0$.

In the scenario we have five types of exotic processes: 
(1) $\mu$-$e$ conversion in a nucleus, 
(2) $\mu^- e^+$ production at LHC, 
(3) dijet production at LHC, 
(4) non-standard interaction of neutrinos,  
(5) muonium conversion. 
In the situation that the $\mu$-$e$ conversion is discovered while 
other cLFV will never be found, we discuss whether 
we can confirm the $\mu$-$e$ conversion signal with the five types 
processes or not. 
Details of each process and the formulation of reaction rates are 
given in Ref.~\cite{RPV}.

\vspace{-3mm}
\section{Numerical Result}  \label{Sec:result} 
\vspace{-2mm}

\begin{figure}[!h]
\hspace{-4mm}
\begin{tabular}{cc}
\subfigure[$\text{N}=\text{Si}$ and $\sqrt{s}= 14$TeV.]{
\includegraphics[scale=0.52]{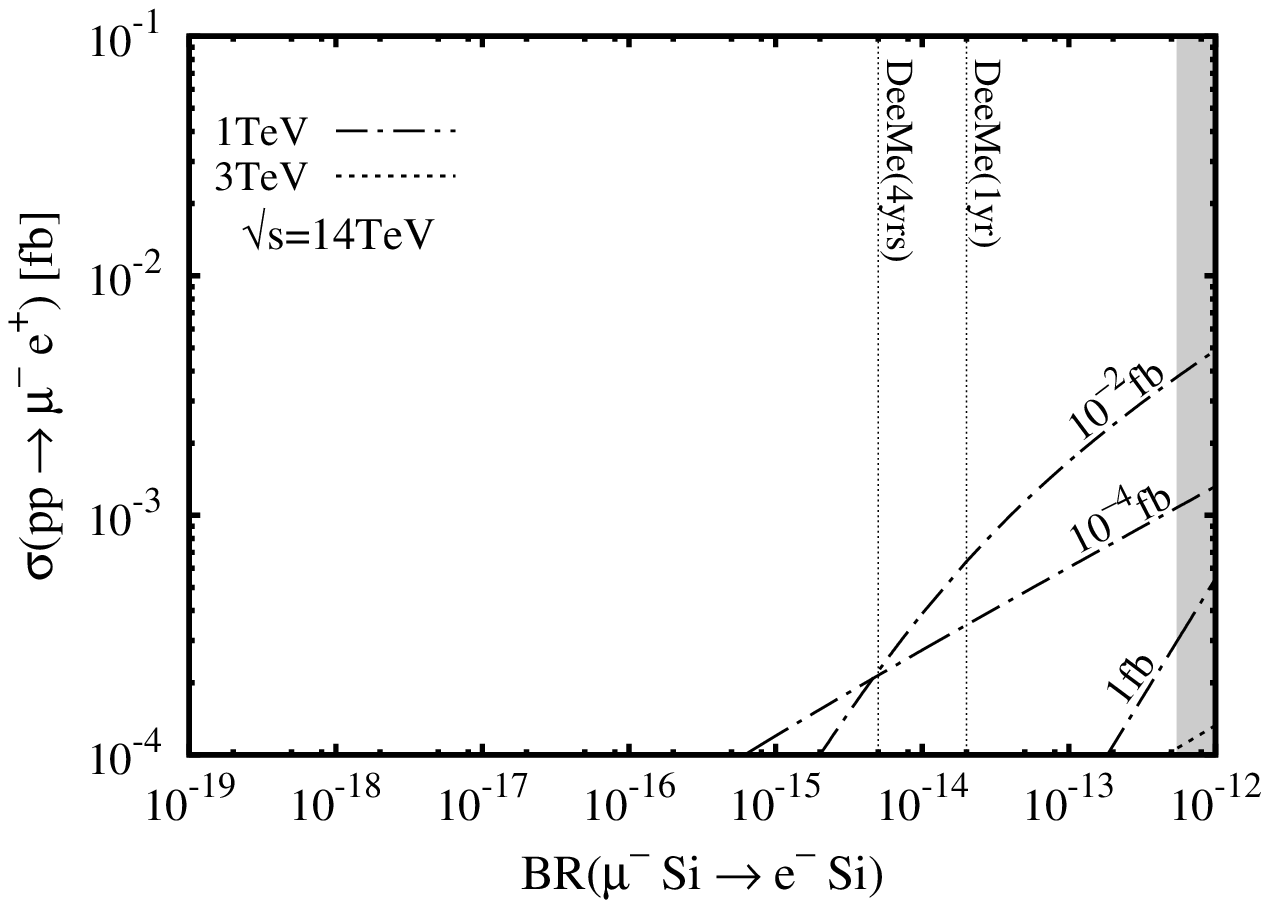}
\label{case1_a}
} & \hspace{-4mm}
\subfigure[$\text{N}=\text{Si}$ and $\sqrt{s}= 100$TeV.]{
\includegraphics[scale=0.52]{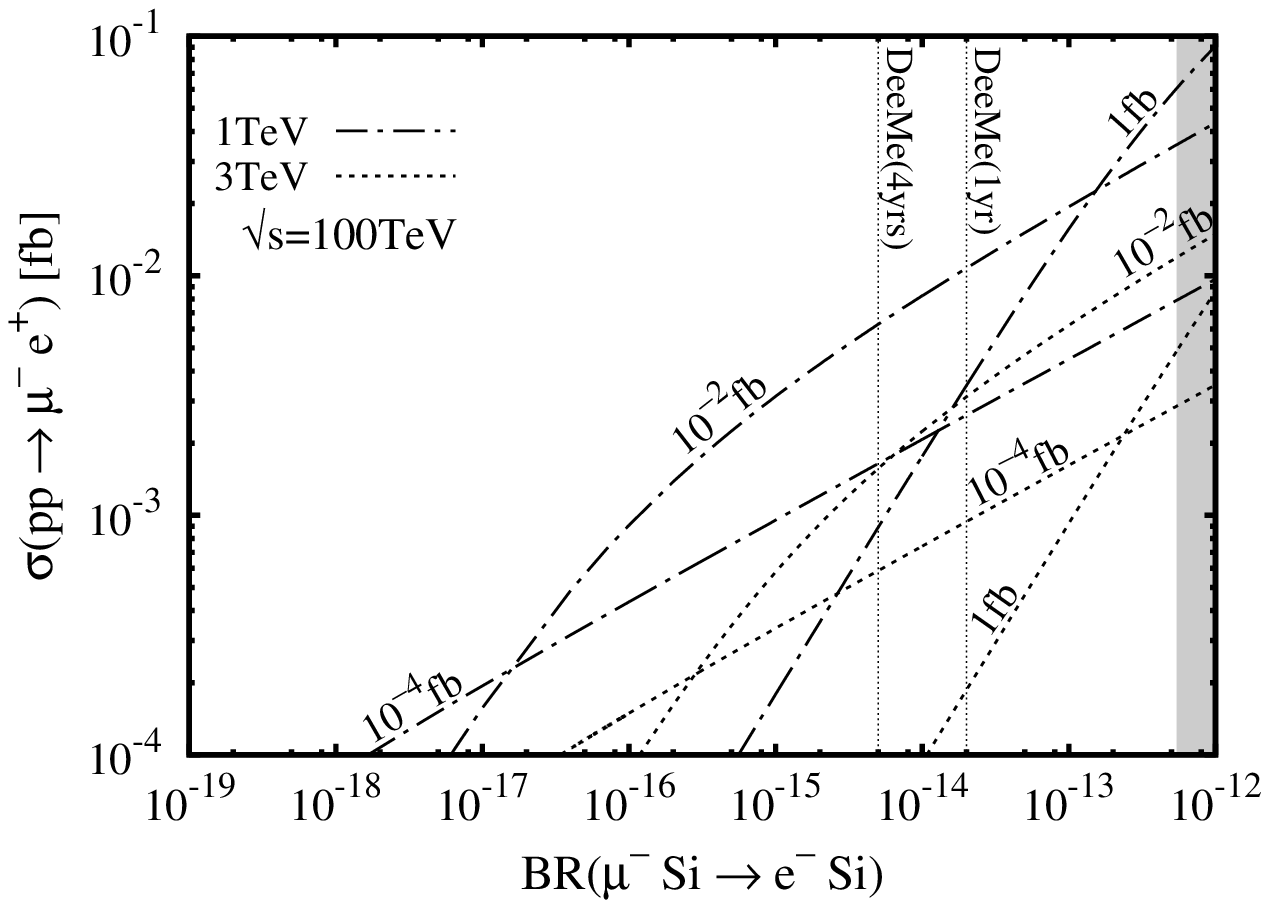}
\label{case1_b}
} \\[-5.5mm]
\subfigure[$\text{N}=\text{Al}$ and $\sqrt{s}= 14$TeV.]{
\includegraphics[scale=0.52]{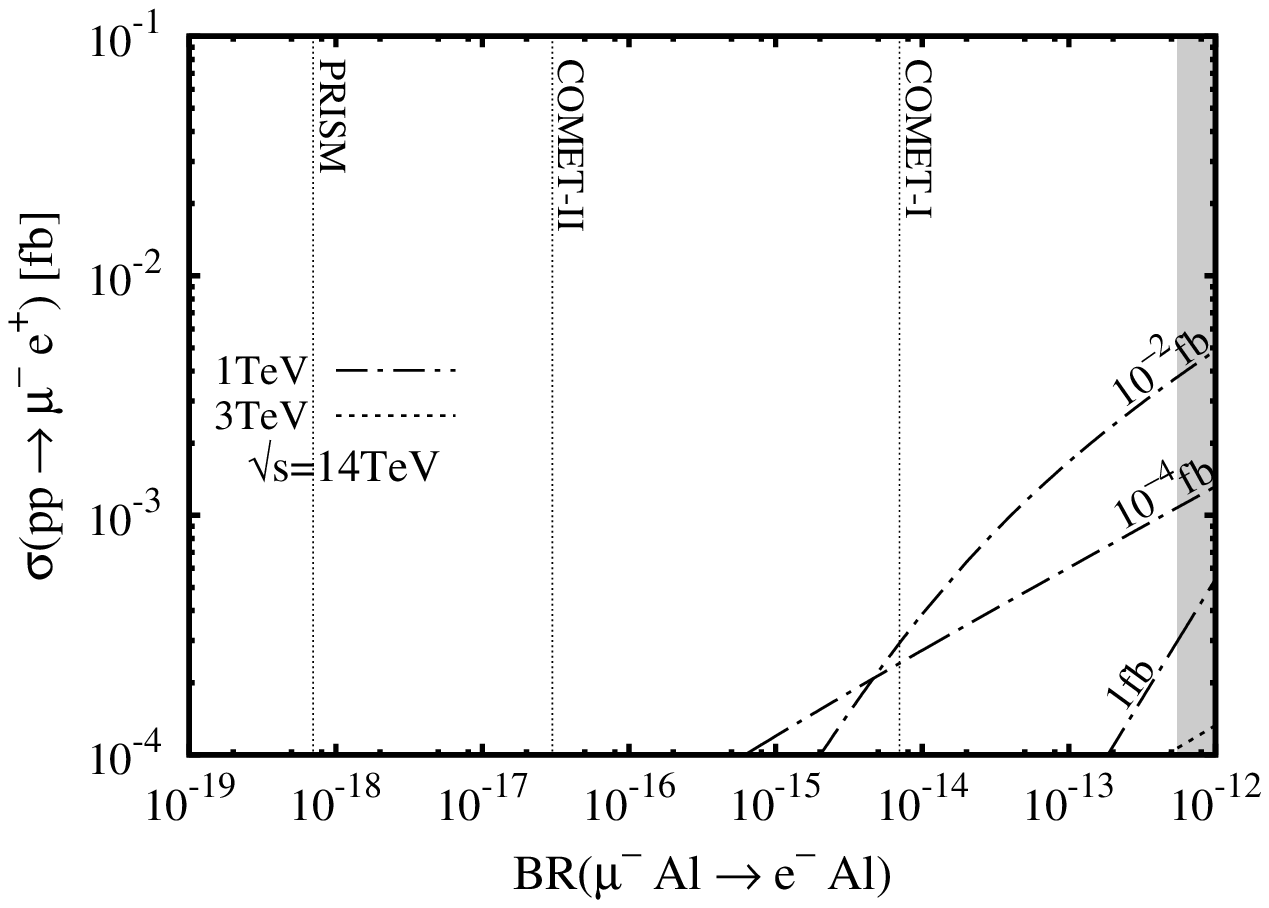}
\label{case1_c}
} & \hspace{-4mm}
\subfigure[$\text{N}=\text{Al}$ and $\sqrt{s}= 100$TeV.]{
\includegraphics[scale=0.52]{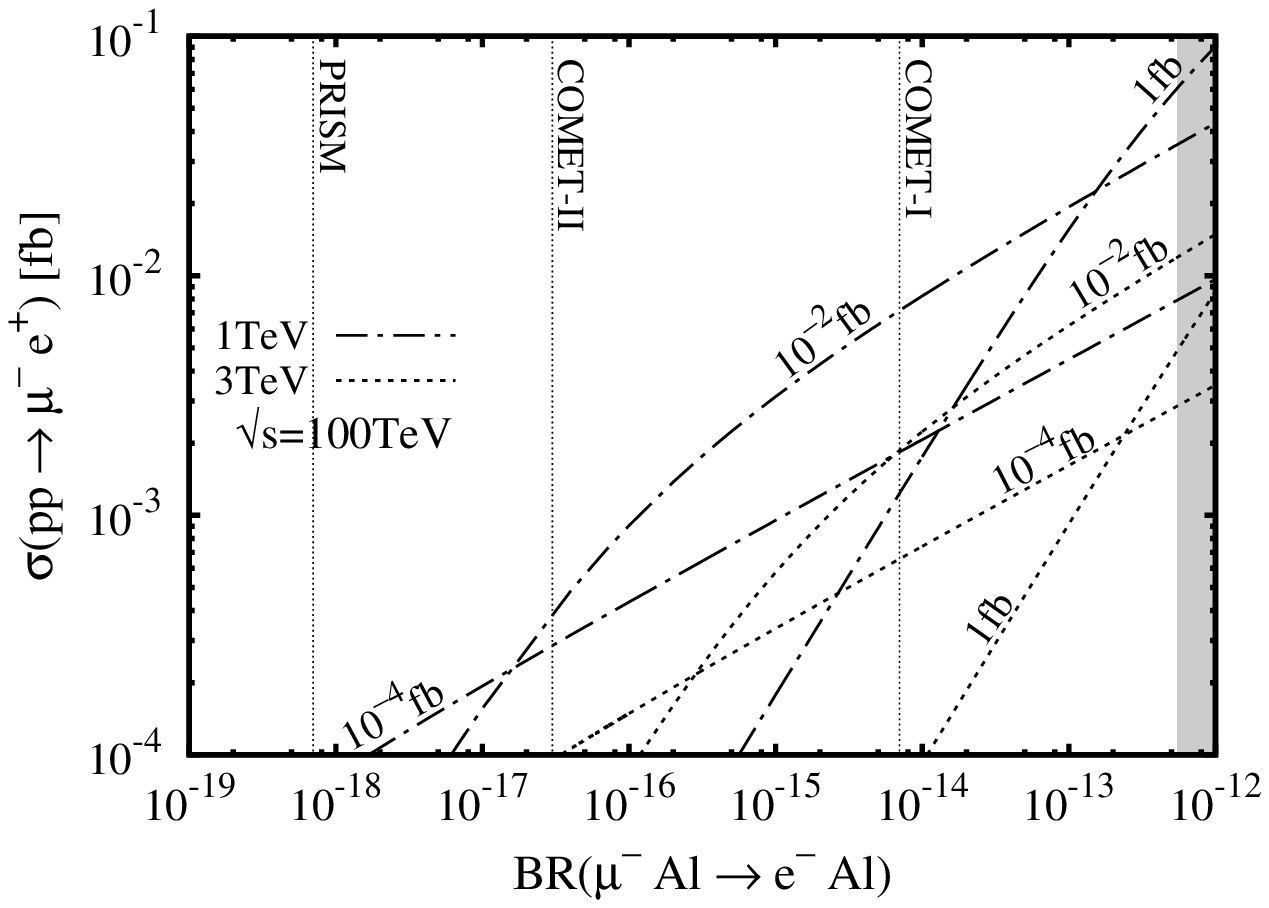}
\label{case1_d}
} \\
\end{tabular}
\caption{$\sigma (pp \to \mu^- e^+)$ as a function of $\text{BR} 
(\mu^- N \to e^- N)$ for each $\sigma (pp \to jj)$ 
in the case-I. $\sigma (pp \to jj)$ are attached on each line. 
Results for $m_{\tilde \nu_\tau} = 1\text{TeV}$ 
($m_{\tilde \nu_\tau} = 3\text{TeV}$) are given by dot-dashed 
(dotted) line. Shaded region is the excluded 
region by the SINDRUM-I\hspace{-1pt}I experiment. Left and right 
panels show the results for $\sqrt{s} = 14\text{TeV}$ and for 
$\sqrt{s} = 100\text{TeV}$, respectively.  We take Si [(a), (b)],  
and Al [(c), (d)] for the target nucleus of $\mu$-$e$ conversion. }
\label{Fig:sigma_vs_BR_I_1}
\end{figure}

The $\mu$-$e$ conversion is a clear signal for 
the RPV scenarios, but it is not the sufficient evidence. 
We must check the correlations among the reaction rates of 
$\mu$-$e$ conversion, the cross sections of $pp \to 
\mu^- e^+$ and $pp \to jj$, and so on to discriminate 
the case-I, -I\hspace{-1pt}I, and -I\hspace{-1pt}I\hspace{-1pt}I, 
and to confirm the scenario. 
Fig.~\ref{Fig:sigma_vs_BR_I_1} shows $\sigma(pp \to \mu \bar e)$ 
as a function of $\text{BR} (\mu + N \to e + N)$ in the case-I. 
Vertical lines show the reach of DeeMe 1-year 
(4-years) running, COMET phase-I (phase-I\hspace{-1pt}I), 
and PRISM. Shaded regions are the excluded region by the 
SINDRUM-I\hspace{-1pt}I~\cite{Bertl:2006up}. 
Each line corresponds to the dijet production cross section, 
$\sigma(pp \to jj)$, at $\sqrt{s}=14\text{TeV}$ (left panels) 
and at $\sqrt{s}=100\text{TeV}$ (right panels), respectively. 
For simplicity, we take universal RPV coupling, $\lambda \equiv 
\lambda_{312} = \lambda_{321} = -\lambda_{132} = 
-\lambda_{231}$. 
Fig.~\ref{Fig:sigma_vs_BR_I_1} shows the clear correlations among 
$\sigma (pp \to \mu^- e^+)$, $\sigma(pp \to jj)$, and $\text{BR} 
(\mu^- N \to e^- N)$. Checking the correlations makes possible to 
distinguish the RPV scenario and other models.

\vspace{-5mm}
\section{Summary and Discussion}  \label{Sec:summary} 
\vspace{-3mm}

We have studied a supersymmetric standard model without R parity
as a benchmark case that COMET/DeeMe observe $\mu - e$ conversion 
prior to all the other experiments observing new physics. 
In this case with the assumption that only the third generation sleptons
contribute to such a process, we need to assume that $\{\lambda'_{311} 
{\rm \hspace{1.2mm} and/or} \hspace{1.2mm} \lambda'_{322}\} \times
\{\lambda_{312}{\rm{ \hspace{1.2mm} and/or \hspace{1.2mm} }}
\lambda_{321}\}$ must be large. 
With the assumptions, we considered the sensitivity of the future 
$\mu$-$e$ conversion experiments on the couplings and slepton masses.
Then with the sensitivity kept into mind
we estimated the reach to the couplings by calculating the cross section
of $pp \rightarrow \mu^- e^+$ and $pp \rightarrow jj$. To have a 
signal of $\mu^- e^+$
both the coupling $\lambda'$ and $\lambda$ must be large and hence
there are lower bounds for them
while to observe dijet event via the slepton only the coupling
$\lambda'$ must be large and hence there is a lower bound on it. 
In all cases we have a chance to get confirmation of $\mu - e $
conversion in LHC.

\vspace{-4mm}


\begin{thebibliography}{99} 
\vspace{-2mm}

\bibitem{RPV}
  J.~Sato and M.~Yamanaka,
  Phys.\ Rev.\ D {\bf 91} (2015) 5,  055018. 




\bibitem{Kuno:1999jp} 
  Y.~Kuno and Y.~Okada: 
  Rev.\ Mod.\ Phys.\  {\bf 73} (2001) 151.

\bibitem{Brooks:1999pu} 
  M.~L.~Brooks {\it et al.}  [MEGA Collaboration]: 
  Phys.\ Rev.\ Lett.\  {\bf 83} (1999) 1521.

\bibitem{Adam:2013mnn} 
  J.~Adam {\it et al.}  [MEG Collaboration]: 
  Phys.\ Rev.\ Lett.\  {\bf 110}, no. 20 (2013) 201801.

\bibitem{Bertl:2006up} 
  W.~H.~Bertl {\it et al.}  [SINDRUM II Collaboration]: 
  Eur.\ Phys.\ J.\ C {\bf 47} (2006) 337.

\bibitem{Bellgardt:1987du} 
  U.~Bellgardt {\it et al.}  [SINDRUM Collaboration]: 
  Nucl.\ Phys.\ B {\bf 299} (1988) 1.



\bibitem{Cui:2009zz} 
  Y.~G.~Cui {\it et al.}  [COMET Collaboration]: 
  KEK-2009-10.


\bibitem{Kuno:2013mha}
  Y.~Kuno [COMET Collaboration]: 
  PTEP {\bf 2013} (2013) 022C01.

\bibitem{Natori:2014yba} 
  H.~Natori [DeeMe Collaboration]: 
  Nucl.\ Phys.\ Proc.\ Suppl.\  {\bf 248-250} (2014) 52.


\bibitem{Hisano:1995cp} 
  J.~Hisano, T.~Moroi, K.~Tobe and M.~Yamaguchi: 
  Phys.\ Rev.\ D {\bf 53} (1996) 2442.

\bibitem{Sato:2000ff} 
  J.~Sato and K.~Tobe: 
  Phys.\ Rev.\ D {\bf 63} (2001) 116010.



\bibitem{Koike:2010xr} 
  M.~Koike, Y.~Kuno, J.~Sato and M.~Yamanaka: 
  Phys.\ Rev.\ Lett.\  {\bf 105} (2010) 121601.




\bibitem{Weinberg:1981wj}
  S.~Weinberg: 
  Phys.\ Rev.\ D {\bf 26} (1982) 287.


\bibitem{Sakai:1981pk}
  N.~Sakai and T.~Yanagida: 
  Nucl.\ Phys.\ B {\bf 197} (1982) 533.


\bibitem{Hall:1983id}
  L.~J.~Hall and M.~Suzuki: 
  Nucl.\ Phys.\ B {\bf 231} (1984) 419.











\end{thebibliography}
\end{document}